# Controlling the helicity of magnetic skyrmions by electrical field in frustrated magnets

Xiaoyan Yao[1*], Jun Chen[1], Shuai Dong[1*]

1 School of Physics, Southeast University, Nanjing 211189, China

* Correspondence author: yaoxiaoyan@seu.edu.cn, sdong@seu.edu.cn.

**Abstract** The skyrmions generated by frustration in centrosymmetric structures host extra internal degrees of freedom—vorticity and helicity, resulting in distinctive properties and potential functionality, which are not shared by the skyrmions stemming from the Dzyaloshinskii-Moriya interaction in noncentrosymmetric structures. The present work indicates that the magnetism-driven electric polarization carried by skyrmions provides a direct handle for tuning helicity. Especially for the in-plane magnetized skyrmions, the helicity can be continuously rotated and exactly picked by applying an external electric field for both skyrmions and antiskyrmions. The in-plane uniaxial anisotropy is beneficial to this manipulation.

# 1 Introduction

In past century, the topological magnetic textures stabilized by complicated interactions were predicted and investigated extensively [1-6]. In particular, topological vortex-like magnetic skyrmions, have attracted a great deal of interest from both the academic and technological fields [7,8]. Since the experimental discovery in 2009 [9], magnetic skyrmions have been observed in many different materials including metals [10-12], semiconductors [13-15], insulators [16,17], and thin film systems [18,19]. Besides the formation of independent skyrmion excitations in the ferromagnetic background, these skyrmions also crystallize into stable lattice, surviving even down to zero-temperature. Experiments revealed that the motion of skyrmion can be driven and controlled by the ultralow electric currents in the metallic system [20,21]. On the other hand, the skyrmions can magnetically induce electric polarization in insulators or semiconductors, which enables the modulation of skyrmions by an external electric field without losses due to Joule heating [22-25]. The magnetoelectric manipulation of skyrmions generated enormous interest in their applications to information storage and processing [26-29].

Up to now, magnetic skyrmions are mostly observed in experiments on noncentrosymmetric materials or interfacial symmetry-breaking heterostructures. It has been widely confirmed that these skyrmions are primarily induced by the Dzyaloshinskii-Moriya interaction (DMI), stemming from inversion-symmetry-breaking and the spin-orbit interaction of different kinds. For example, Bloch-type skyrmions can be stabilized in chiral magnets with the Dresselhaus spin-orbit interaction, while Néel-type skyrmions may appear in polar magnets with the Rashba spin-orbit interaction or in magnetic ultrathin films with interfacial DMI [30-33]. Meanwhile, the magnetic antiskyrmions discovered in Heusler compounds are stabilized by the anisotropic DMI with opposite signs along two orthogonal in-plane directions [34-38]. Although different skyrmions may be observed in different materials, their morphology in one compound is always fixed because the DMI is determined by the structure. On the other hand, various alternative mechanisms are proposed, especially for the skyrmions in centrosymmetric materials, such as the long-ranged magnetic dipolar interactions [17], the four-spin exchange interactions [18], the coupling between itinerant electron spins and localized spins [39,40]. In particular, the magnetic frustration has long served as a relatively simple yet rich source of novel magnetic phases [41-45]. In 2012, it was first theoretically proposed that the skyrmion crystal can be stabilized by the frustrated interactions [46], but the following experimental reports on the skyrmions in frustrated systems were very limited [47,48]. It was exciting that very recently the skyrmion crystal was experimentally discovered in the frustrated centrosymmetric triangular-lattice magnet $Gd_2PdSi_3$ [49] and breathing kagomé lattice magnet $Gd_3Ru_4Al_{12}$ [50]. Due to the frustration source, the skyrmions are typically much smaller than the DMI-driven counterpart, and thus contribute to a giant topological Hall effect. Later, nanometric square skyrmion lattice was reported in a centrosymmetric tetragonal magnet $GdRu_2Si_2$ [51].

It has been theoretically predicted that the frustration-induced skyrmions in centrosymmetric structure show distinctive properties [52-56]. One fascinating advantage is the extra internal degrees of freedom—vorticity and helicity. The helicity-dependent current responses unveiled by the dynamics simulation imply that these factors not only characterize the morphology of skyrmion, but also can be exploited to control its motion and furthermore to achieve diverse functionality. On the contrary, for the DMI system, both vorticity and helicity are locked by the sign and direction of the DMI vector. The only way to tune the helicity of the DMI-driven skyrmion is the controllable DMI, i.e. to tune the spin-orbit coupling [57,58]. But the helicity is still locked for a certain composition in these previous reports. Compare to the fixed morphology of the DMI-driven skyrmion, the frustration-induced skyrmion holds the intrinsic degrees of freedom to tune its morphology, and therefore shows the related exotic behaviors not shared by the DMI systems. How to take advantage of these freedom degrees to realize an effective and flexible manipulation remains a crucial physical problem to be resolved. Very recently it was reported that the helicity of dipolarly stabilized skyrmions can be tuned by varying the material parameters and geometry of nanostructure [59,60].

In this paper, the frustration-induced magnetic skyrmions in both crystal and individual forms are investigated on a triangular lattice model with competing exchange interactions. For the conventional skyrmions in the perpendicularly magnetized case, the noncollinear spin configuration may generate an electric polarization normal to the lattice plane with its value depending on helicity. Since the energies of skyrmions with different helicities are degenerate, a low electric field may flip the helicity between 0 and π through a continuous variation. More accurate tunability can be achieved in the unconventional skyrmions in the in-plane magnetized case. The nonzero electric polarization with magnitude and direction depending on helicity emerges for both skyrmions and antiskyrmions. An electric field of constant magnitude rotating in the lattice plane may rotate the helicity. If the uniaxial anisotropy is in the lattice plane and in the same direction of magnetic field, this rotation is continuous and the helicity can be exactly picked by changing the direction of electric field. When the uniaxial anisotropy is normal to the lattice plane, the degeneracy of helicity is lifted partially and six symmetric values of helicity are preferred.

## 2  Model and methods

The frustration-driven skyrmion crystal was first predicated as metastable state realized in triangular lattice under applied magnetic field perpendicular to the lattice plane at moderate temperatures [46]. The later investigation testified that an easy-axis anisotropy perpendicular to the lattice plane is sufficient to stabilize the skyrmion crystal with the lowest energy at zero-temperature [53,54,61]. Here the frustrated magnetic model on two-dimensional triangular lattice is considered with the Hamiltonian expressed as

$$H = -J_1 \sum_{<i,j>} S_i \cdot S_j - J_2 \sum_{<<i,m>>} S_i \cdot S_m - J_3 \sum_{<<<i,n>>>} S_i \cdot S_n - k\sum_i (S_i \cdot e_k)^2 - h\sum_i S_i \cdot e_h \quad (1)$$

where $S_i$ represents a classic spin of unit length at the $i$-th site on $xy$-plane. The first three terms on the right of Eq. 1 are the exchange energies, where $J_1$, $J_2$ and $J_3$ represent the exchange interactions between spins on the nearest-neighboring (<i,j>), the second-nearest-neighboring ($<<i, m>>$) and the third-nearest-neighboring ($<<<i, n>>>$) sites. Ferromagnetic $J_1=1$ is fixed as the energy unit, and all the parameters are simplified with reduced units. It has been reported that either of antiferromagnetic $J_2$ and $J_3$ considered is enough to produce a stable skyrmion crystal, and the similar results can be obtained in these two cases [46,61]. So $J_2=0$ and $J_3=-0.5$ are set unless otherwise noted. The fourth and fifth terms describe the energies of uniaxial anisotropy and magnetic field, where the uniaxial anisotropy with the magnitude $k$ along $e_k$ direction and the magnetic field with the magnitude $h$ along $e_h$ direction.

Due to the frustration source, the present skyrmion is typically much smaller than the DMI-driven counterpart. The noncollinear spin texture in such a scale of the lattice constant usually provides the condition to produce electric polarization according to the spin current mechanism [62], where an electric dipole $p_{ij}$ can be induced by two neighboring canting spins ($S_i$ and $S_j$) in the form of

$$p_{ij} = -e_{ij} \times (S_i \times S_j) \tag{2}$$

where $e_{ij}$ denotes the unit vector connecting the two sites of neighboring $S_i$ and $S_j$. The total electric polarization ($P$) is estimated by the summary over all the bonds, which also can be written as the summary of all the local electric polarization on site ($p_i$), namely

$$P = \sum_{[i,j]} p_{ij} = \sum_i p_i. \tag{3}$$

Considering the ferroelectric energy with external electric field ($E$), the total Hamiltonian can be written as

$$H = -J_1 \sum_{<i,j>} S_i \cdot S_j - J_3 \sum_{<<<i,n>>>} S_i \cdot S_n - k \sum_i (S_i \cdot e_k)^2 - h \sum_i S_i \cdot e_h - E \cdot P. \tag{4}$$

The simulation is performed on the triangular lattice of sites $N=5184$ with periodic boundary conditions. Different lattice sizes are checked to confirm the stability of the main results. The Metropolis algorithm combined with the over-relaxation method is applied to find the stable state with the lowest energy [63,64]. On every parameter point, the system is first evolved from a relatively high temperature to a very low temperature gradually, and then the energy is further minimized to approach the limit of zero temperature. The final result is obtained by comparing independent data sets evolving from different initial states. The spin dynamics at zero temperature is discussed by using the fourth-order Runge-Kutta method to numerically solve Landau-Lifshitz-Gilbert (LLG) equation as below.

$$\frac{\partial S}{\partial t} = -S \times h_{eff} + \alpha S \times \frac{\partial S}{\partial t} \tag{5}$$

where the Gilbert damping coefficient $\alpha=0.2$ to ensure quick relaxation to the equilibrium state, and the time $t$ is measured in units of $\hbar/J_1$. $h_{eff} = -\frac{\partial H}{\partial S}$ is the effective field with Hamiltonian $H$ defined in Eq. 4.

The obtained spin configuration is characterized by the spin structure factor, which is evaluated for three spin components ($\gamma=x$, $y$ and $z$) respectively as follows,

$$S^\gamma(q) = \sum_{i,j} e^{iq\cdot(r_j-r_i)} \langle S_i^\gamma \cdot S_j^\gamma \rangle \tag{6}$$

The topological character is confirmed by the skyrmion number or topological charge ($Q$), which is defined as

$$Q = \frac{1}{4\pi} \iint S \cdot (\partial_x S \times \partial_y S) \, dxdy . \tag{7}$$

$Q$ quantifies the number of times spin vectors wrapping around a unit sphere as the coordinate ($x$, $y$) spans the whole planar space, which can be calculated for a spin lattice in the manner as Ref. [65]. To elucidate the detailed nature, the local topological charge density $\rho(r)$ can be expressed as

$$\rho(r) = \frac{1}{4\pi} S \cdot (\partial_x S \times \partial_y S) . \tag{8}$$

For an isolated skyrmion in the ferromagnetic background in two-dimensional system with a perpendicular magnetic field, the morphology is determined by vorticity and helicity. If the polar coordinate ($r$, $\varphi$) is used with the symmetric center of spin texture as the origin, the spin vector can be expressed as $S=(\sin\theta\cos\phi, \sin\theta\sin\phi, \cos\theta)$. $\theta$ only depends on $r$. $\phi=n\varphi+\eta$, where integer $n$ is the vorticity and $\eta$ is the helicity. Skyrmion and antiskyrmion are distinguished conventionally by the sign of $n$, i.e. $n=1$ for skyrmion and $n=-1$ for antiskyrmion [7]. The helicity has continuous value from $-\pi$ to $\pi$, where $\eta=-\pi$ and $\pi$ are identical. In particular, when $n=+1$, $\eta=0$ or $\pi$ is the Néel type, $\eta=-\pi/2$ or $\pi/2$ is the Bloch type. All the structures of antiskyrmions ($n$=-1) are equivalent on rotation in the $xy$-plane [7].

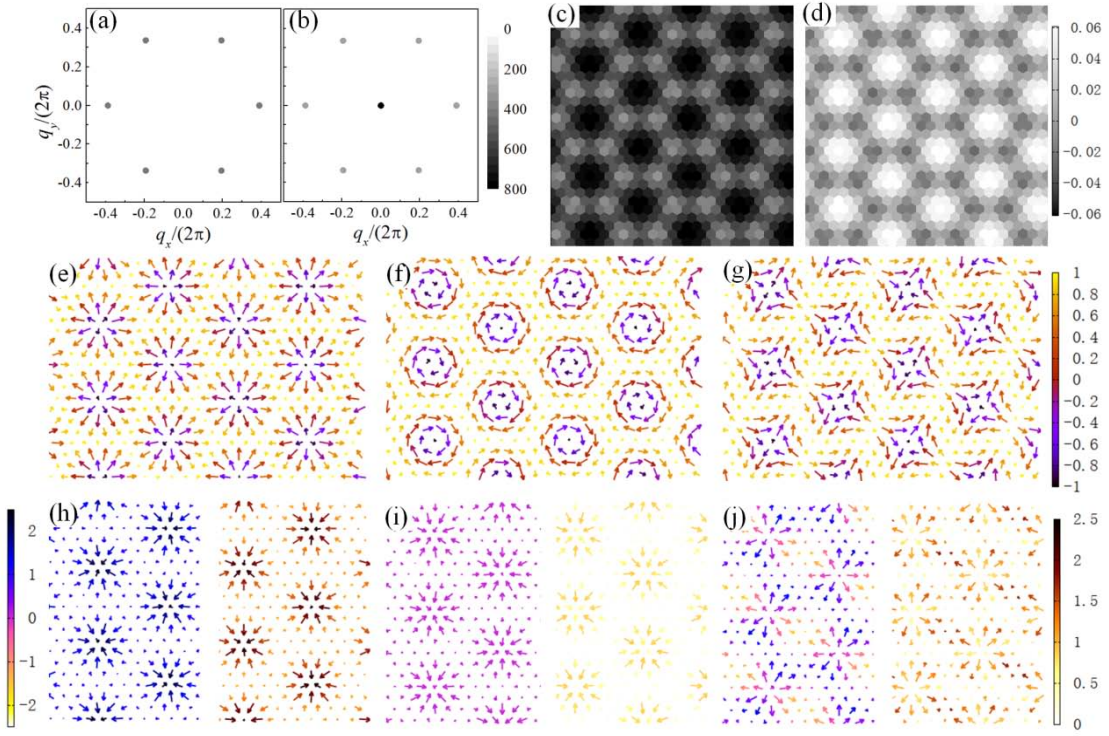

**Fig. 1** The stable PM (anti)skyrmion crystals with the anisotropy $k$=0.1 and the magnetic field $h$=0.3 (both $\bm{e}_k$ and $\bm{e}_h$ along +$z$ direction). The intensity plots of the spin structure factor: (a) the sum of $x$ and $y$ components $S^x(\bm{q})$+$S^y(\bm{q})$ and (b) $z$ component $S^z(\bm{q})$. The gray scale represents intensity. The maps of topological charge density $\rho(\bm{r})$ for (c) skyrmion crystal and (d) antiskyrmion crystal, where the gray scale represents the value of $\rho(\bm{r})$. The real-space spin configurations of three typical skyrmion crystals: (e) Néel skyrmion crystal with $n$=+1 and $\eta$=0, (f) Bloch skyrmion crystal with $n$=+1 and $\eta$=-$\pi$/2, (g) the antiskyrmion crystal with $n$=-1 and $\eta$=0.66$\pi$. The vectors show the projections of spins onto $xy$-plane, and the color refers to the $z$ components of spins. (h)-(j) The corresponding maps of the local electric polarization on site ($\bm{p}_i$). The vectors show the projections of $\bm{p}_i$ onto $xy$-plane. The color refers to the $z$ components of $\bm{p}_i$ in the left half and the magnitude |$\bm{p}_i$| in the right half. For visibility, only part of the lattice is plotted.

## 3  Results and discussion

For the frustrated triangular model with $J_1$-$J_2$ or $J_1$-$J_3$, rich phase diagrams have been reported [53,54,61,66]. Under the consideration of uniaxial anisotropy and magnetic field both perpendicular to the lattice plane ($\bm{e}_k$ and $\bm{e}_h$ along +$z$ direction), the perpendicularly magnetized (PM) skyrmion crystal emerges at intermediate $h$ and above a small $k$, and remains down to zero-temperature. Consistent with the previous reports, the spin configuration of skyrmion crystal phase in the present simulation shows the typical spin structure factor of a triple-$q$ magnetic orderings [46], as presented in Figs. 1(a) and 1(b). The topological character is confirmed by the nonzero value of topological charge, and the corresponding topological charge density $\rho(\boldsymbol{r})$ demonstrates a triangular lattice for skyrmions and antiskyrmions respectively as plotted in Figs. 1(c) and 1(d). Due to the inversion symmetry in the frustrated system, the skyrmion crystals with different vorticities and helicities are energetically degenerate. But for one state in the skyrmion crystal phase with the lowest energy, the symmetry is spontaneously broken, and thus one vorticity and one helicity are selected arbitrarily. As shown in Figs. 1(e-g), each spin configuration is composed of skyrmions with the uniform vorticity and helicity, and thus $Ns=|Q|$ represents the number of skyrmions.

Although all these skyrmion crystals are energetically degenerate, they show electric polarization $\boldsymbol{P}$ depending on the helicity, which provides a handle for electric field to tune helicity. For skyrmions of $n=1$, the direction of $\boldsymbol{P}$ is always along $z$ axis, namely all the $xy$-components of $\boldsymbol{p}_i$ cancel out each other, as illustrated in Figs. 1(h, i). For Néel skyrmion of $\eta=0$ ($\eta=\pi$), all $z$-components of $\boldsymbol{p}_i$ show positive (negative) values, while they give values near zero for Bloch type of $\eta=\pm\pi/2$. In the case of antiskyrmion ($n=-1$) as plotted in Fig. 1(j), all the components of $\boldsymbol{p}_i$ cancel out each other, i.e. $\boldsymbol{P}=0$. The $\boldsymbol{P}$ dependence on $\eta$ for skyrmion of $n=1$ is plotted in Fig. 2(a). The magnitude of $\boldsymbol{P}$ ($P$) varies with $\eta$ continuously, following $P=P_m\cos(\eta)$ where $P_m$ is the amplitude. Thus $P$ could be an characteristic index of $\eta$, which may be measured more easily. Néel skyrmion crystal with $\eta=0$ or $\pi$ always presents the maximum value of $P$ ($P_m$) with opposite orientations. For the Néel skyrmion crystal, increasing $h$ or $k$ reduces $P_m$ owing to the suppression of the noncollinear spin parts. In addition, $Ns$ mainly depends on the frustration tuned by the exchange interaction $J_3$. When the frustration is enhanced by raising $|J_3|$ with $h$ and $k$ fixed, $Ns$ increases and $P_m$ also, but $P_m$ per skyrmion ($P_m/Ns$) decreases as plotted in Figs. 2(b-d).

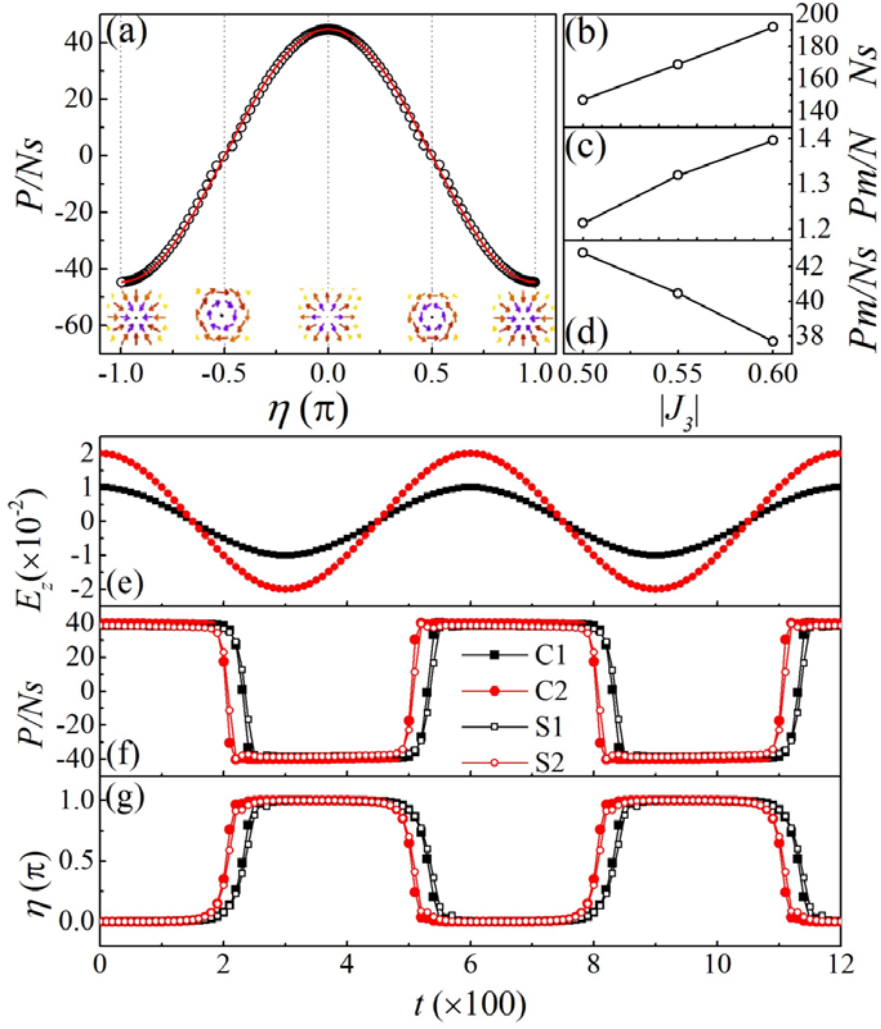

**Fig. 2** The static properties without $E$ (a-d) and the dynamical behaviors with $E$ (e-g) for the PM skyrmions with $e_k$ and $e_h$ both along +z direction. (a) $P$ per skyrmion ($P/Ns$) as a function of $\eta$ for the stable PM skyrmion crystals at $k=0.1$ and $h=0.3$. The red solid curve shows $P=P_m\cos(\eta)$, and the insets at the bottom illustrate the typical skyrmions at the positions marked by the dotted lines. (b) The number of skyrmions $Ns$, (c) $P_m$ per site ($P_m/N$) and (d) $P_m$ per skyrmion ($P_m/Ns$) as functions of $|J_3|$ with $k=0.1$ and $h=0.3$ fixed. (e) AC electrical field along $z$-axis $E_z=E_m\cos(2\pi t/600)$ with different amplitude $E_m=0.01$ and $0.02$, is applied on metastable skyrmion crystal and isolated skyrmion at the same parameter point of $k=0.4$ and $h=0.5$. The time dependences of $P$ and $\eta$ are plotted in (f) and (g), where C1: skyrmion crystal with $E_m=0.01$; C2: skyrmion crystal with $E_m=0.02$; S1: isolated skyrmion with $E_m=0.01$; S2: isolated skyrmion with $E_m=0.02$.

Since the states with different $\eta$ are energetically degenerate, it is easy to align skyrmions with different $\eta$ to Néel type of $\eta=0$ or $\pi$ by applying a low electric field. Although the spatially homogeneous electric field cannot move the skyrmion [67], $P$ can be flipped by $E$ without energy loss. When an AC electrical field $E_z=E_m\cos(2\pi t/600)$ with $E_m=0.01$ and $0.02$ is applied along $z$-axis (Fig. 2(e)), $P$ flips between $P_m$ and $-P_m$, and simultaneously $\eta$ flips between $0$ and $\pi$, as illustrated in Figs. 2(f) and 2(g). If $E_m$ is stronger, the flips occur more promptly. The independent PM skyrmion also exists in the ferromagnetic state near the skyrmion crystal phase in this system. The simulation on the metastable isolated PM skyrmion shows the very similar magnetoelectric behavior. For instance, an isolated skyrmion presents nearly the same $E$-driven flips of $P$ and $\eta$ as plotted in Figs. 2(f) and 2(g), where the same parameters are adopted for crystal and isolated states for a better comparison.

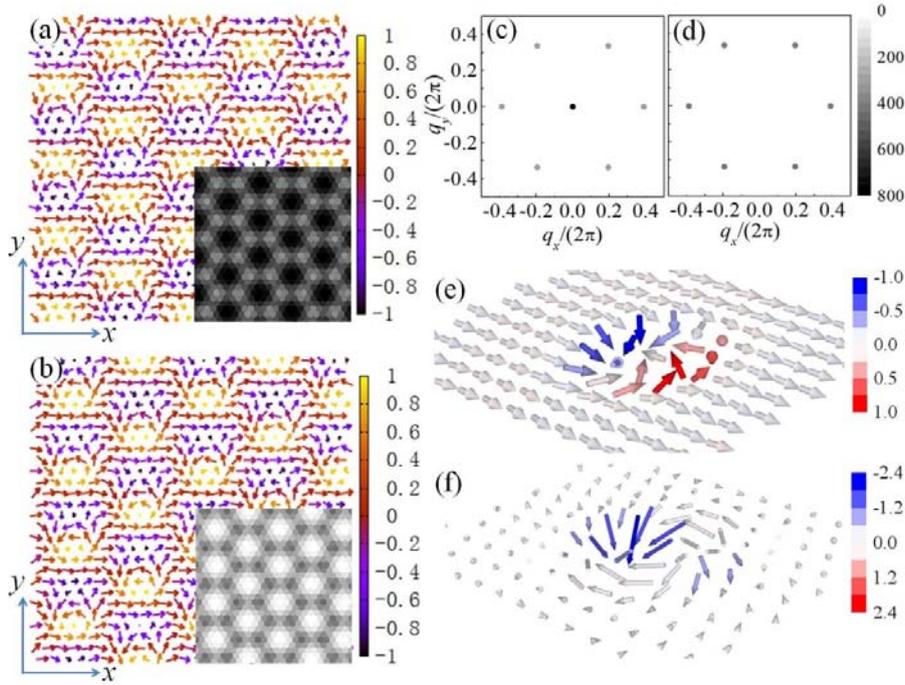

**Fig. 3** The spin configurations of IM (anti)skyrmions. The stable IM crystals of (a) skyrmions ($n=+1$ and $\eta=\pi/2$) and (b) antiskyrmions ($n=-1$ and $\eta=\pi/2$) with the anisotropy $k=0.1$ and the magnetic field $h=0.3$ (both $e_k$ and $e_h$ along $+x$ direction). The arrows show the projections of spins onto $xy$-plane and the color refers to the $z$-components. The corresponding $\rho(r)$ maps are inserted in the lower right corners where the color represents the value of $\rho(r)$ as shown in Fig. 1(d). The intensity plots of the spin structure factor: (c) $S^x(q)$ and (d) $S^y(q)+S^z(q)$, where the gray scale represents the intensity. (e) The metastable isolated IM skyrmion of $\eta=\pi$ in the FM background at $k=0.4$ and $h=0.5$ (both $e_k$ and $e_h$ along $+x$ direction). (f) The corresponding map of $p_i$. The color refers to the $z$ components for both (e) and (f). For visibility, only part of the lattice is plotted.

It is noteworthy that the conventional skyrmions in most investigations are the PM skyrmions with the spin at the core antiparallel to the spins at the perimeter both perpendicular to the lattice plane. As discussed above, the tunability on helicity by electrical field is limited for the PM skyrmions. It is hard to exactly pick one certain helicity for skyrmion, and it is nearly impossible to control the helicity for antiskyrmion without $P$. It is worth noting that a novel in-plane magnetized (IM) skyrmion, where the spin at the core still antiparallel to the spins at the perimeter but both within the lattice plane, was put forward in 2019 [68,69]. This unconventional skyrmion was also known as one kind of bimeron [70], and the similar spin texture had been observed in experiments [71]. It is very convenient to produce the IM skyrmions in the present frustrated model. According to the Hamiltonian in Eq. 4, applying magnetic field in the same direction of anisotropy, i.e. $e_k=e_h$, along z-axis or one in-plane direction, the scenario and the phase diagram is the same. In particular, when both $e_k$ and $e_h$ along +x direction, the IM skyrmions can form stable triangular lattice surviving down to zero temperature (Figs. 3(a) and 3(b)). The spin structure factor also demonstrates the typical triple-$q$ feature, only except that $x$-component swaps place with $z$-component as shown in Figs. 3(c) and 3(d). The isolated IM skyrmion can also exit in the ferromagnetic state near the skyrmion crystal phase as plotted in Fig. 3(e).

The IM skyrmion can be obtained from the PM skyrmion by a 90° rotation around the $y$-axis [68], which can be realized by rotating magnetic field. The IM skyrmion is topologically equivalent to the corresponding PM one with the same $Q$, because they can be associated with a smooth deformation of the spin texture. Therefore, the topological properties are well retained in this IM version of skyrmion, and a pure topological Hall effect was predicted [68]. The two internal degrees of freedom—vorticity and helicity are also retained, and thus they can be estimated as the PM case.

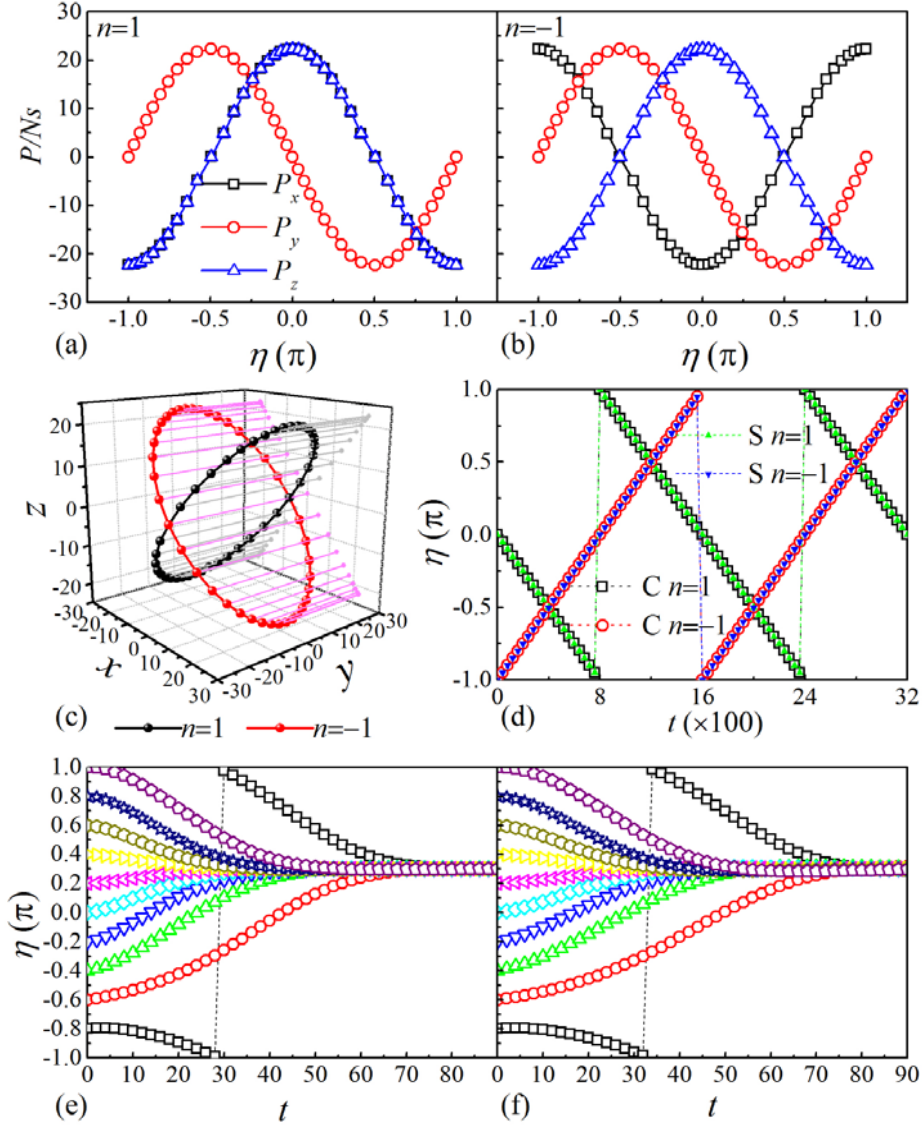

**Fig. 4** The static properties without $E$ (a-c) and the dynamical behaviors with $E$ (d-f) for the IM (anti)skyrmions with $e_k$ and $e_h$ both along $+x$ direction. The components of $P/Ns$, namely $P_x$, $P_y$ and $P_z$ as functions of $\eta$ for (a) skyrmion and (b) antiskyrmion crystals in the stable IM state of $k=0.1$ and $h=0.3$. (c) The corresponding three-dimensional trajectories of $P/Ns$ for both skyrmion ($n=1$) and antiskyrmions ($n=-1$) crystals with the projection on $xz$-plane along [101] and [10$\bar{1}$] respectively. For the metastable IM states at $k=0.4$ and $h=0.5$, when $E$ of the fixed magnitude $E_m=0.01$ is rotated in $xy$-plane by $\varphi_E=0.00125\pi t$, the time dependences of $\eta$ are plotted in (d) where C: skyrmion crystal and S: isolated skyrmion. Helicity $\eta$ as functions of $t$ from varied initial $\eta$ to reach $\eta=0.3\pi$ by applying $E$ of (e) $E_x=0.01\cos(-0.3\pi)$ and $E_y=0.01\sin(-0.3\pi)$ for IM skyrmion crystal, (f) $E_x=0.01\cos(0.3\pi+\pi)$ and $E_y=0.01\sin(0.3\pi+\pi)$ for IM isolated antiskyrmion.

The IM skyrmion can be regarded as a pair of tight-binding meron and antimeron as plotted in Fig. 3(e). The corresponding $p_i$ map demonstrates that the IM skyrmion produces the local electric polarization in an asymmetric way. It is attractive that in this case with $e_k$ and $e_h$ both along $x$-axis, $P$ shows nonzero value for all the (anti)skyrmions of different $\eta$. Furthermore, both orientation and magnitude of $P$ depend on vorticity and helicity. Figs. 4(a) and 4(b) plot the components of $P$ per skyrmion as functions of $\eta$, presenting different behaviors for skyrmion and antiskyrmion crystals. $P_x=P_z=P_a\cos(-\eta)=P_a\cos(\eta)$ and $P_y=P_a\sin(-\eta)=-P_a\sin(\eta)$ in the skyrmion case, while $P_x=P_a\cos(\eta+\pi)=-P_a\cos(\eta)$, $P_y=P_a\sin(\eta+\pi)=-P_a\sin(\eta)$ and $P_z=-P_a\cos(\eta+\pi)=P_a\cos(\eta)$ in the antiskyrmion case. That is, when $\eta$ is changed, $P$ rotates along an ellipse on the plane normal to $[10\bar{1}]$ ($[101]$) when $n=1$ ($n=-1$), as illustrated in Fig. 4(c). The similar behavior can be observed for the isolated IM (anti)skyrmion. Therefore, $P$ provides a handle for $E$ to tune $\eta$. A small $E$ with constant magnitude applied in the $xy$-plane can efficiently manipulate the helicity for skyrmion and antiskyrmion respectively. Here $\varphi_E$ is used to denote the angle from $+x$ to $E$ direction in $xy$-plane, If $E$ of the magnitude $E_m=0.01$ is rotated in $xy$-plane by $\varphi_E=0.00125\pi t$, then the different linear variations of $\eta$ is exhibited in Fig. 4(d) for skyrmions and antiskyrmions respectively, i.e. $\eta=(2\pi)-\varphi_E$ for $n=1$ and $\eta=\varphi_E-\pi$ for $n=-1$. The crystal and isolated states show nearly the same variation when the same parameters are adopted. The linear correspondence provides a simple way to pick one $\eta$ exactly and flexibly, which is effective to both skyrmion and antiskyrmion in both crystal and isolated forms. For example, to reach $\eta=0.3$ for $n=1$, $E$ of $E_x=0.01\cos(-0.3\pi)$ and $E_y=0.01\sin(-0.3\pi)$ can be applied, while in the case of $n=-1$, that would be $E_x=0.01\cos(0.3\pi+\pi)$ and $E_y=0.01\sin(0.3\pi+\pi)$. Whatever value for the original $\eta$, the system can reach $\eta=0.3\pi$ soon, as plotted in Figs. 4(e) and 4(f).

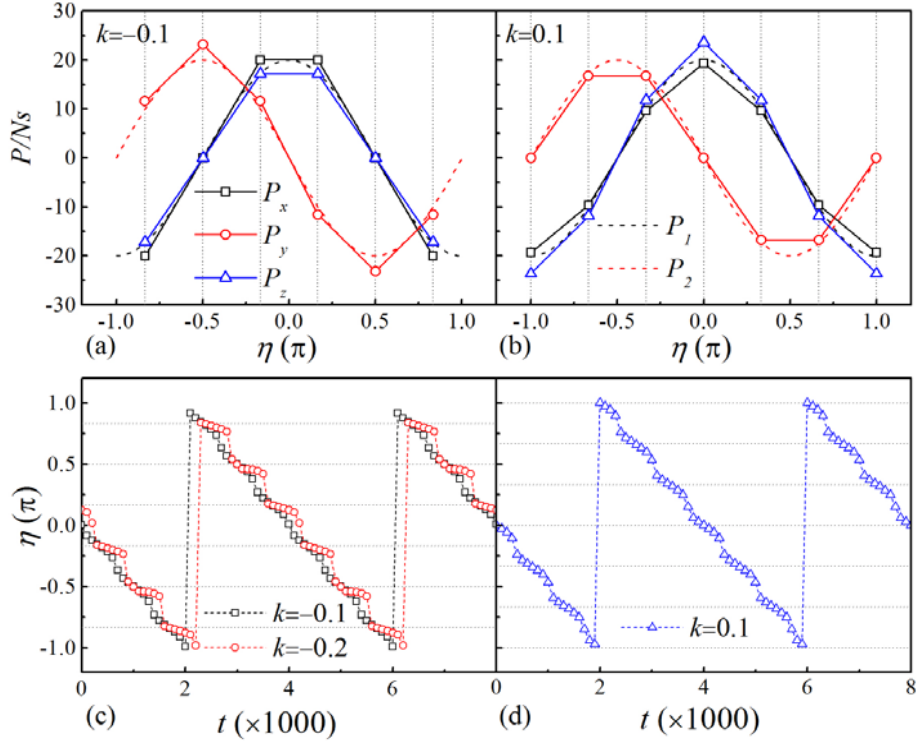

**Fig. 5** The static properties without $E$ (a,b) and the dynamical behaviors with $E$ (c,d) for the IM skyrmion crystals under the magnetic field of $h=0.4$ and $e_h$ along $+x$ direction and the anisotropy of $e_k$ along $+z$ direction. The components of $P/Ns$, namely $P_x$, $P_y$ and $P_z$ as functions of $\eta$ for (a) stable state of $k=-0.1$ and (b) metastable state of $k=0.1$. Dashed lines of $P_1 = P_a \cos(\eta)$ and $P_2 = -P_a \sin(\eta)$ with $P_a=20$ are plotted as guides to eye. When $E$ with the fixed magnitude $E_m=0.01$ is rotated in $xy$-plane by $\varphi_E=0.0005\pi t$, the time dependences of $\eta$ are plotted for (c) $k=-0.1$ and $-0.2$, (d) $k=0.1$. The gray dotted lines mark the six $\eta$ values mentioned in the text.

It should be mentioned that the IM skyrmion crystal as stable state down to zero temperature can also be realized with easy-plane anisotropy ($k<0$) along $z$-axis and in-plane magnetic field. When $k$ is non-negligible, different from the continuous $\eta$ mentioned above, six $\eta$ values are preferred in this scenario, i.e. $\eta = -\frac{5\pi}{6}, -\frac{\pi}{2}, -\frac{\pi}{6}, \frac{\pi}{6}, \frac{\pi}{2}, \frac{5\pi}{6}$, as displayed in Fig. 5(a). Interestingly, other six values i.e. $\eta = -\frac{2\pi}{3}, -\frac{\pi}{3}, 0, \frac{\pi}{3}, \frac{2\pi}{3}, \pi$, will be preferred in the metastable IM skyrmion crystal with easy-axis anisotropy ($k>0$) along $z$-axis and in-plane magnetic field, as plotted in Fig. 5(b). It is means that the degeneracy of helicity is resolved partially by the nonzero $k$, and six states with lower energy remain. The corresponding $P$ still roughly follows the rules mentioned above, except that $P_x$ and $P_z$ do not overlap any more due to *k* along *z*. When ***E*** with the magnitude $E_m$=0.01 is rotated by $\varphi_E$=0.0005πt in *xy*-plane, the time-dependence of $\eta$ demonstrates the sloping steps corresponding to the six $\eta$ values, respectively for *k*<0 and *k*>0 as displayed in Figs. 5(c) and 5(d). The stronger anisotropy induces more obvious steps, implying more robust states with the six $\eta$ values. In this case, the skyrmions are actually deformed by anisotropy. However, as long as the anisotropy is not so strong, the skyrmion crystal can be well preserved due to the topological protection. On the other hand, the isolated skyrmion may also appear. But it is hard to produce isolated skyrmion in ferromagnetic background in this parameter region. Therefore it is hard to achieve an efficient electric control of $\eta$ because of the complex background.

It should be mentioned that the IM skyrmions in both crystal and isolated forms may survive as metastable state in the case without anisotropy (*k*=0). Here the helicity can be tuned continuously and exactly for skyrmion and antiskyrmion crystals respectively, just like the case with in-plane uniaxial anisotropy (Fig. 4), whereas the isolated IM skyrmion can not be controlled freely also due to the complex background. The better tunability can be achieved for crystal form because the contribution from the background is relatively small.

Although the magnetic dipole interaction prefers the Bloch-type texture, the corresponding energy cost of helicity variation is very small, and it is even much smaller for IM skyrmions than that of PM skyrmions. The electric field applied will overcome this energy barrier to realize the helicity control. (See Supplemental Material Note 1 for details.) The magnetic skyrmion is a natural topological texture in two spacial dimensions. In three dimensional space, the situation will be complicated and interesting for frustrated interlayer interaction, where the characterization and tenability of internal freedom degrees will be explored in the future studies (See Supplemental Material Note 2 for details.)

## 4 Conclusions

The PM and IM skyrmions driven by frustration are investigated on a triangular lattice with competing interactions in both crystal and isolated forms. Aiming at insulating systems, the magnetoelectric properties are focused on to explore unique electric controllability on the intrinsic helicity degree of freedom. The magnetism-driven electric polarization provides a direct handle for the external electrical field to enable the manipulation of helicity with the topological charge conserved. The flip of $\eta$ between 0 and $\pi$ accompanied by the reversible ***P*** is observed for the PM skyrmions. The more intriguing tunability can be achieved for the IM skyrmions, where the different dependences of ***P*** on $\eta$ are revealed for skyrmions and antiskyrmions respectively. Take this advantage, the helicity can be continuously rotated and exactly picked by applying ***E*** in both skyrmion and antiskyrmion cases. The in-plane uniaxial anisotropy favors this manipulation, whereas the anisotropy perpendicular to the lattice plane weakens the continuity of the tunability. In addition, the crystal form can be controlled easier than the isolated one due to less contribution from the background.

**Acknowledgments**

This work is supported by the research grants from the National Natural Science Foundation of China (Grant Nos. 11834002 and 11674055).

# Supplementary Materials for

# Controlling the helicity of magnetic skyrmions by electrical field in frustrated magnets

Xiaoyan Yao[1*], Jun Chen[1], Shuai Dong[1*]

1 School of Physics, Southeast University, Nanjing 211189, China

* Correspondence author: yaoxiaoyan@seu.edu.cn, sdong@seu.edu.cn.

**Supplementary Note 1:**

It has been well known that the magnetic dipole interaction (MDI) contributes to micron-sized skyrmion bubbles [1]. Although MDI prefers the Bloch-type texture and removes the degeneration of helicity, it is very small comparing with exchange interaction. Therefore, the energy cost of helicity variation due to MDI is very small, which can be overcome by electric field to realize the helicity control. The threshold of the electric field can be estimated roughly by $\frac{\mu_0 \mu_s^2}{4\pi a^3 P_s}$ where $a$ is the lattice constant, $\mu_0$ is the permeability of vacuum, and $\mu_s$ is the atomic magnetic moment, and $P_s$ is the average electric polarization on site driven by noncollinear spin texture. It is worth noting that the in-plane magnetized skyrmions are more preferred by MDI, and the energy cost and the electric field threshold of helicity variation will be much smaller than those of the perpendicularly magnetized skyrmions.

**Supplementary Note 2:**

In bulk, along the magnetic field direction, a magnetic skyrmion typically extends in the manner of a tube or a line [2]. The simulation on three dimensional frustrated centrosymmetric magnets indicated that a crystal of such vertical skyrmion lines is stabilized in the case of nearest-neighbor ferromagnetic interlayer exchange interaction, while frustration of the interlayer interactions leads to multiple ways of skyrmion stacking [3]. Typical skyrmion straight line in the case of uniform ferromagnetic interlayer interaction presents the same skyrmion with the same helicity on every layer. The situation will be complicated and interesting for multiple stacking skyrmions in the case of frustrated interlayer interactions. This is beyond the scope of this manuscript, but will be explored in the future studies.

**Supplementary References**